# Analysis of Peculiarities of the SEHRS Spectrum of $4,4^{'}$ - Bipyridine Molecule on the Base of the Dipole-Quadrupole Theory

**V.P. Chelibanov[1] A.V. Golovin[2] A.M. Polubotko[3]**

[1]State University of Information Technologies, Mechanics and Optics, Kronverkskii 49, 197101 Saint Petersburg, RUSSIA  E-mail: Chelibanov@gmail.com

[2]Saint Petersburg State University, Ulianovskaya st. 1, 198504, Saint Petersburg Petrodvoretz RUSSIA E-mail: golovin50@mail.ru

[3] A.F. Ioffe Physico-Technical Institute, Politechnicheskaya 26, 194021 Saint Petersburg, RUSSIA E-mail: alex.marina@mail.ioffe.ru

## Abstract

The SEHRS spectrum of $4,4^{'}$- Bipyridine is analyzed on the base of the Dipole-Quadrupole theory. It is demonstrated that there appear strong lines caused by vibrations transforming after a unit irreducible representation of the $D_2$ symmetry group, which is most probably describe the symmetry properties of the molecule. These lines are nearly forbidden for the molecule, adsorbed on rough metal surfaces. Appearance of these lines is associated with a strong quadrupole light – molecule interaction, which exists in this system. In addition, there are lines caused by contributions from both the vibrations transforming after the unit irreducible representation $A$ and the representation $B_1$, which describe transformational properties of the $d_z$ component of the dipole moment, which is perpendicular to the surface. This result is associated with the specific geometry of the molecule, when the indicated vibrations can be nearly degenerated and cannot be resolved by the SEHRS spectra analysis. Analysis of the SEHRS spectra for the possible geometry of the molecule with the $D_{2h}$ symmetry group leads to similar results. This issue is in a full coincidence with the results of the SEHRS Dipole-Quadrupole theory.

# Introduction

Surface Enhanced Optical processes are of great interest, since these processes can be a powerful tool for scientific investigations and are applicable in physics, chemistry and biology. Usually one explains these phenomena by the hypothesis of surface plasmons and a chemical enhancement. However, these approaches do not explain appearance of forbidden lines, which are observed in the spectra of all these processes in molecules with sufficiently high symmetry. This difficulty is overcome in the Dipole-Quadrupole theory, which explains appearance of these lines naturally due to existence of a strong quadrupole light – molecule interaction, arising due to strongly inhomogeneous fields, which exist near a rough metal surface[1]. In SEHRS the forbidden lines associated with the totally symmetric vibrations, transforming after the unit irreducible representation were observed in pyrazine and phenazine. In addition, such lines must be observed in other symmetric molecules with sufficiently high symmetry that follows from the results of the Dipole-Quadrupole SEHRS theory[2]. However, there are no experimental data on the SEHRS spectra of other symmetric molecules, which could confirm appearance of this type of forbidden lines. In this work we present theoretical interpretation of the SEHRS spectrum of $4,4^{'}$ - Bipyridine, observed in[3]. The problem is that the symmetry group of the molecule may be under the question and can be $D_2$ or $D_{2h}$. However, it was demonstrated in[4] that this molecule most probably belongs to the $D_2$ symmetry group. This result was obtained on the base of the energy minimum fundamental principle for this configuration. Here we shall demonstrate that the most enhanced lines are caused by vibrations transforming respectively either by the unit irreducible representation, or by the representation $B_1$, which describes transformational properties of the dipole moment component $d_z$, which is perpendicular to the surface or by both of them. The first one is associated with the strong quadrupole light – molecule interaction, when the strongest scattering arises due to the quadrupole moments $Q_{xx}, Q_{yy}, Q_{zz}$, or their linear combinations, having a constant sign, which are named as main quadrupole moments. The second one is associated with the scattering via the two main quadrupole moments and due to the strong dipole interaction with the $d_z$ moment, associated with the enhancement of the $E_z$ component of the electric field, which is perpendicular to the surface. However, the most of the enhanced lines are associated with both enhancement mechanisms since both types of vibrations are nearly degenerate and contribute to the same lines. Here we mean that the lines, caused by these nearly degenerate vibrations cannot be resolved in fact in the above mentioned experiment[3]. This unique situation strongly differs from the one for pyrazine and phenazine and is associated with a specific geometry of the molecule, which consists of two weekly interacting benzene rings and, therefore, there may exist two nearly degenerated symmetric and antisymmetric vibrational states with very close frequencies.

## Main relations of the Dipole-Quadrupole SEHRS theory

The Dipole-Quadrupole SEHRS theory was published in[2]. In addition, there is the Dipole-Quadrupole theory of SEIRA and SERS, which is based on the same conceptions. The last one is published in the monograph[1]. Therefore, we expound here only some main topics of this theory and send the reader for detailed acquaintance to the above works.

In accordance with the Dipole-Quadrupole theory, there exists a strong quadrupole light – molecule interaction arising in surface fields strongly varying in space near rough metal surface. This interaction is associated with the quadrupole terms in the light – molecule interaction Hamiltonian for the incident and scattered fields

$$\widehat{H}_{e-r}^{\text{inc}} = \left| \overline{E}_{inc} \right| \frac{(\overline{e}^* \overline{f}_e^*)_{inc} e^{i\omega_{inc}t} + (\overline{e}\,\overline{f}_e)_{inc} e^{-i\omega_{inc}t}}{2} \quad , \tag{1}$$

$$\widehat{H}_{e-r}^{\text{scat}} = \left| \overline{E}_{scat} \right| \frac{(\overline{e}^* \overline{f}_e^*)_{scat} e^{i\omega_{scat}t} + (\overline{e}\,\overline{f}_e)_{scat} e^{-i\omega_{scat}t}}{2} \tag{2}$$

Here $\bar{E}_{inc}$ and $\bar{E}_{scat}$ are the incident and scattered electric fields, $\omega_{inc}$ and $\omega_{scat}$ are corresponding frequencies, $\bar{e}$ is a corresponding polarization vector,

$$f_{e\alpha} = d_{e\alpha} + \frac{1}{2E_\alpha} \sum_\beta \frac{\partial E_\alpha}{\partial x_\beta} Q_{e\alpha\beta} \qquad (3)$$

is an $\alpha$ component of generalized vector of interaction of light with the electrons of the molecule and

$$d_{e\alpha} = \sum_i ex_{i\alpha} \text{ and } Q_{e\alpha\beta} = \sum_i ex_{i\alpha} x_{i\beta}, \qquad (4)$$

are an $\alpha$ component of the dipole moment vector and the $\alpha\beta$ component of the quadrupole moments tensor of electrons. Here $x_{i\alpha}$ and $x_{i\beta}$ are the Cartesian coordinates of the $i$ th electron. In accordance with the Dipole – Quadrupole theory, only the terms of the Hamiltonians (1, 2), which are associated with the moments $Q_{xx}, Q_{yy}, Q_{zz}$ and $d_z$ are essential for the strong scattering. These moments are named as main moments $Q_{main}$ and $d_{main}$. Corresponding reasoning one can find in our publications and in[1], for example. The Hamiltonian with the quadrupole and dipole moments can be very strong due to some features of the scattering on these quadrupole moments: enhancement of the $E_z$ component of the electric field, which is perpendicular to the surface and anomalously high values of derivatives $\frac{\partial E_\alpha}{\partial x_\alpha}$ with equal indices. Because of the quadrupole interaction important role in the SEHRS process, various combinations of the dipole and quadrupole moments can contribute to the effect (Figure 1). Then the scattering is determined by several contributions, which arise due to the scattering via their various combinations. Thus the scattering is determined by these contributions, which occur due to these various types of the scattering and the SEHRS cross-section in symmetrical molecules is proportional to the sum of them[2]

$$\sigma_{SEHRS_{(s,p)}} \sim \left| \sum_{f_1, f_2, f_3} T_{(s,p), f_1-f_2-f_3} \right|^2 dO \qquad (5)$$

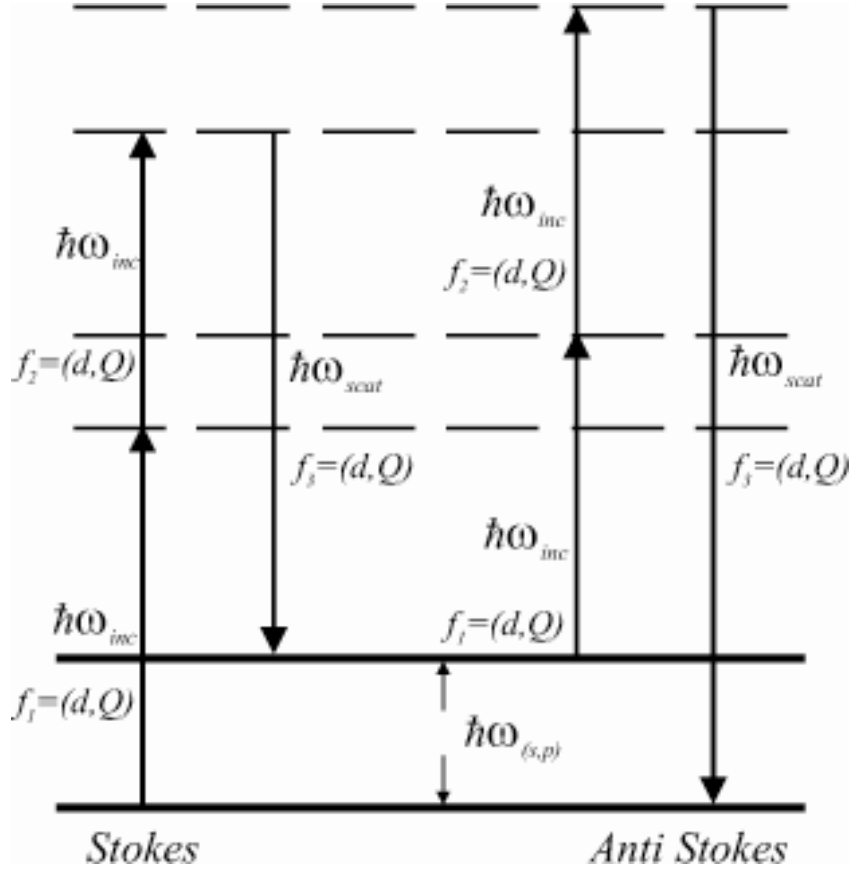

Figure 1. The scattering diagrams of SEHRS for the Stokes and Anti Stokes scattering. The virtual absorption and emission can occur via various dipole and quadrupole moments $f_1, f_2$ and $f_3$.

Here under $f_1, f_2$ and $f_3$ we mean various dipole and quadrupole moments, $T_{(s,p),f_1-f_2-f_3}$ means the contribution, which arise to the scattering via various moments. The indices $(s, p)$ refer to degenerate vibrational modes, where the index $s$ numerates the group of degenerated vibrational states, while $p$ numerates the states inside the group. The contributions $T_{(s,p),f_1-f_2-f_3}$, which we will simply designate further as $(f_1 - f_2 - f_3)$, obey selection rules

$$\Gamma_{(s,p)} \in \Gamma_{f_1} \Gamma_{f_2} \Gamma_{f_3} \quad , \tag{6}$$

where the symbol $\Gamma$ designates the irreducible representation, which describes transformational properties of the $(s, p)$ vibration and the $f_1, f_2$ and $f_3$ moments. There may be several distinct contributions to one line, however the magnitude of these contributions can significantly differ one from another. Therefore, we can sometimes take into account only one, the most enhanced contribution that really defines the scattering cross-section value. Since the quadrupole interaction with the main quadrupole moments can be most enhanced, the most enhanced contributions are those, which describe the scattering via three main quadrupole moments: $(Q_{main} - Q_{main} - Q_{main})$. Because the electric field component $E_z$, which is perpendicular to the surface is strongly enhanced too, the contribution $(Q_{main} - Q_{main} - d_z)$ will be enhanced as well, but with a lesser degree. The other enhanced contributions $(Q_{main} - d_z - d_z)$ and $(d_z - d_z - d_z)$ will be enhanced too but with a lesser degree than the previous two. Since the main quadrupole moments transform after the unit irreducible representation, the most enhanced lines are caused by the vibrations with the unit irreducible representation and with the representation, which describes transformational properties of the $d_z$ moment, which is perpendicular to the surface in accordance with the selection rules (6). The lines caused by vibrations transforming after the unit irreducible representation are forbidden in usual Hyper Raman scattering in molecules with sufficiently high symmetry. This situation is realized in the symmetry groups where the moment $d_z$ transforms after another irreducible representation than the unit one. Such lines were observed in pyrazine and phenazine, which belong to the $D_{2h}$ group[5-7], and satisfies to this condition.

## Analysis of the SEHRS spectrum of 4,4'-Bipyridine

Let us next consider the regularities of the SEHRS spectrum of 4,4'-Bipyridine. The SEHRS spectrum of this molecule was published in[4]. The problem is that the geometry of this molecule is under a question. The authors of [3] determined the geometry which consists from two connected benzene rings, with the connection between carbon atoms and where the opposite carbon and hydrogen atoms are substituted by nitrogen atoms. They came to conclusion that the benzene rings do not lie in the same plane and are turned with respect to each other on the angle 38.7 degrees (Figure 2). The symmetry group of the molecule in this case is $D_2$.

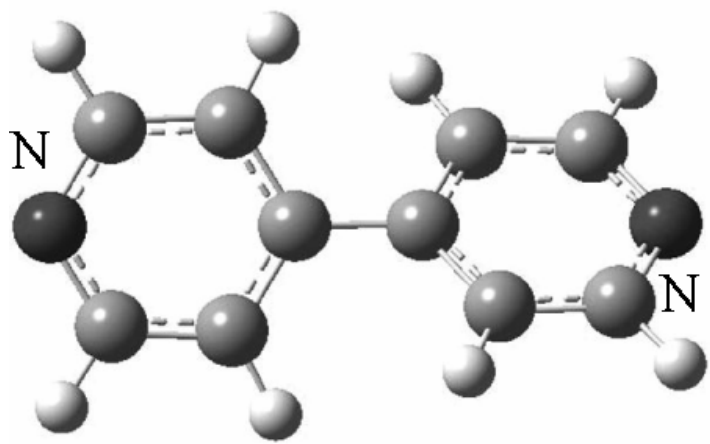

Figure 2. The structure of the 4, 4' Bipyridine molecule. The benzene rings are turned with respect to each other on the angle 38.7 degree.

We made the full geometry optimization and determined the frequencies and symmetry of vibrations of the 4, 4' Bipyridine molecule using the hybrid functional B3LYP with the 6-31G(d) basic set in the Gaussian 03 program[9]. We would like to stress that our calculations with the program Gaussian 03 result in the same conclusion. However in literature researchers often consider that the molecule has a plain geometry and belongs to the $D_{2h}$ symmetry group. Here we consider that the molecule belongs to the $D_2$ symmetry group, which has four irreducible representations. For correct determination of the symmetry of vibrations we need to know orientation of the molecule with respect to the coordinate system. It was chosen is such a manner that the $z$ axis passes via two nitrogen atoms. Since 4, 4'-Bipyridine adsorbs vertically and is connected with the substrate via the nitrogen atom, then the $z$ axis coincides with the enhanced $E_z$ component of the electric field, which is perpendicular to the surface. We calculated vibration wavenumbers and determined the symmetry of vibrations of the molecule by the program Gaussiana 03. The SEHRS spectrum of 4,4' Bipyridine is presented on Figure 3 and the calculated wavenumbers and the assignment are collected in Table 1. In the first column the experimental wavenumbers of the vibrational spectrum taken from[4] and the qualitative estimation of the intensity of spectral lines are presented. The second and the third column present the calculated vibrational wavenumbers, which are most close to the experimental values and the most probable assignment of these vibrations. One can see that the most probable vibrations, which determine the lines of the spectrum are those with the unit irreducible representation $A$ and the irreducible representation $B_1$, which determine transformational properties of the $d_z$ moment. Moreover, they have very close calculated values of wavenumbers and follow in pairs for most part of vibrational modes. The corresponding experimental values of wavenumbers for these vibrational modes are 386, 1008, 1064, 1218, 1484 and 1602 $cm^{-1}$. This effect is associated with the specific geometry of the molecule. Two benzene rings are connected with the $C-C$ bond, which changes the vibrational frequencies of the benzene ring only slightly. Therefore, there are two vibrational states, symmetric and antisymmetric, which must have close vibrational frequencies and various symmetry $A$ and $B_1$. Because of very close values of the wavenumbers with the difference, which can not be resolved in the experiments, both vibrations contribute in the same line in fact and therefore, can not be distinguished from experiment unambiguously. The only line, caused by only one vibration with $A$ symmetry is $1298 cm^{-1}$. Existence of this line points out the existence of the strong quadrupole light – molecule interaction and we can consider that this interaction contributes significantly in the above mentioned lines too. The lines caused by the totally symmetric vibrations with the unit irreducible representation are forbidden in usual HRS in molecules with sufficient high symmetry. However, in molecules with $D_2$ symmetry group they are formally allowed. This situation occurs since there is a contribution in the lines with the unit irreducible representation caused by the scattering via $d_x, d_y$ and $d_z$ moments (the $(d_x - d_y - d_z)$ contribution). However, this contribution can not be strong, but really is very week and must be close to zero since it is determined by tangential components of the electric field $E_x$ and $E_y$, which are parallel to the surface and are nearly equal to zero because of a nearly ideal conductivity of the metal surface. This result is valid for the contributions from the molecules, which are adsorbed in the first layer. However it is well known that the molecules, adsorbed in the second and other layers enhance the SEHRS signal significantly lower than those in the first layer (The first layer effect[8]). Therefore, the above contribution is determined mainly by the scattering from the first layer and must be very week. However, the contribution $(Q_{main} - Q_{main} - Q_{main})$ is huge and determines appearance of the strong lines with the unit irreducible representation $A$. In accordance with the Dipole-Quadrupole theory, appearance of the strong lines with the $B_1$ symmetry is associated with the $(Q_{main} - Q_{main} - d_z)$ types of the scattering. Thus analysis of the SEHRS spectrum of 4,4'- Bipyridine for the above geometry with the $D_2$ symmetry group demonstrates that there is a line at $1298 cm^{-1}$ with $A$ symmetry, which is nearly forbidden in usual HRS. In addition there are the lines, caused both by vibrations with the irreducible representations $A$ and $B_1$. Besides, there are several lines in the spectra, which may be caused by vibrations with the $B_1$ symmetry only. They are the lines with the wavenumbers 628 and $1044 cm^{-1}$. However, in any case the SEHRS spectrum for the $D_2$ symmetry of the molecule is well explained by the Dipole-Quadrupole theory.

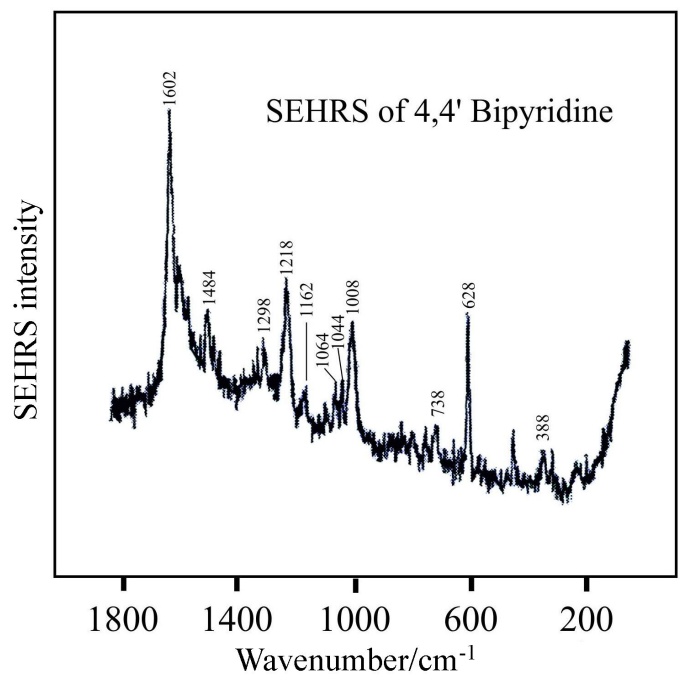

Figure 3. The SEHRS spectrum of 4,4' Bipyridine[3]

Table 1. Assignment of the SEHRS lines of 4, 4'-Bipyridine for $D_2$ symmetry group

| Wavenumber $cm^{-1}$ experiment | The most close calculated values of wavenumbers in $cm^{-1}$ and the most possible assignment. | | Additional possible assignment in the $D_2$ symmetry group | |
|---|---|---|---|---|
| | A | $B_1$ | $B_2$ | $B_3$ |
| 386 w | A (383) | $B_1$ (377) | | |
| 628 s | | $B_1$ (597) | | $B_3$ (658) |
| 736 w | A (743) | | $B_2$ (743) | |
| 1008 m | A (985) | $B_1$ (1029) | | |
| 1044 w | | $B_1$ (1029) | | |
| 1064 w | A (1073) | $B_1$ (1067) | | |
| 1162 w | | absent | | |
| 1218 m | A (1222) | $B_1$ (1218) | | |
| 1298 w | A (1282) | | | |
| 1484 w | A (1507) | $B_1$ (1487) | | |
| 1602 s | A (1605) | $B_1$ (1599) | | |

Table 2. Assignment of vibrations of 4,4' Bipyridine for the planar geometry of the molecule and the $D_{2h}$ symmetry group.

| Experimental wavenumbers $cm^{-1}$ | Calculated wavenumbers $cm^{-1}$ and the most possible assignment. | | |
|---|---|---|---|
| | $A_g$ | $B_{1u}$ | $B_{1g}$ |
| 386 w | | | $B_{1g}$ (379) |
| 628 s | | $B_{1u}$ (598) | |
| 736 w | $A_g$ (742) | | |
| 1008 m | $A_g$ (990) | $B_{1u}$ (1026) | |
| 1044 w | | $B_{1u}$ (1026) | |
| 1064 w | | $B_{1u}$ (1063) | |
| 1162 w | | absent | |
| 1218 m | $A_g$ (1240) | $B_{1u}$ (1222) | |
| 1298 w | $A_g$ (1268) | | |
| 1484 w | | $B_{1u}$ (1489) | |
| 1602 s | $A_g$ (1603) | $B_{1u}$ (1602) | |

Since there is an opinion, that 4, 4' Bipyridine belongs to the $D_{2h}$ symmetry group we made calculations of the vibrational wavenumbers and determination of the symmetry of vibrations for this case too (Table 2). As it was demonstrated by calculations, there are strong lines, caused by vibrations with the unit irreducible representation $A_g$, which are forbidden in usual HRS. In addition, there are strong lines with the $B_{1u}$ irreducible representation, which describe transformational properties of the $d_z$ moment. Sometimes there is no unambiguous assignment for the vibrational modes due to the close values of the calculated wavenumbers of vibrations with $A_g$ and $B_{1u}$ symmetry, which follow in pairs. However, existence of the lines with these irreducible representations demonstrates that even for the plane geometry (which is apparently less probable) the SEHRS spectrum of 4, 4' Bipyridine can be described by the Dipole-Quadrupole theory.

## Conclusion

Thus, our analysis of the SEHRS spectrum of 4, 4-Bipyridine demonstrates existence of sufficiently strong lines, caused by totally symmetric vibrations, transforming after the unit irreducible representation in both possible geometries of the molecule, belonging to the $D_2$ and $D_{2h}$ symmetry groups. Because of very small value of the contribution of the $(d_x - d_y - d_z)$ type, which contribute in the lines with the $A$ symmetry (in a molecule with the $D_2$ symmetry group), we can say that appearance of these lines is a manifestation of the strong quadrupole light-molecule interaction in this system. There are several lines caused by vibrations with the $B_1$ irreducible representation, which indicate existence of the strong dipole light-molecule interaction associated with the enhancement of the $E_z$ component of the electric field, which is perpendicular to the surface. In addition, there are several lines, which can be caused both by the nearly degenerated vibrations with $A$ and $B_1$ symmetry, which apparently can not be resolved in the experiment. The intensity of these lines can be formed both by the strong $(Q_{main} - Q_{main} - Q_{main})$ and $(Q_{main} - Q_{main} - d_z)$ types of the scattering. The peculiarity of the SEHRS spectra with respect to the one for the pyrazine molecule[6,7] is the fact that several lines must be assigned to both irreducible representations $A$ and $B_1$. This situation occur due to specific geometry of the molecule, which consists of two benzene rings, which slightly affect on the vibrations of each other and form symmetric and antisymmetric vibrational states with very close frequencies Investigation of the 4, 4'-Bipyridine molecule with the geometry with the $D_{2h}$ symmetry group leads to the same results.